\documentclass{jpp}
\usepackage{graphicx}
\usepackage{epstopdf, epsfig}
\usepackage{hyperref}
\usepackage{amsmath}
\usepackage{xcolor}

\shorttitle{Measures of quasisymmetry for stellarators}
\shortauthor{E. Rodriguez, E. Paul and A. Bhattacharjee}

\title{Measures of quasisymmetry for stellarators}

\author{E. Rodr\'{i}guez\aff{1,2}
  \corresp{\email{eduardor@princeton.edu}},
  E. Paul\aff{1,2}
 \and A. Bhattacharjee\aff{1,2}}

\affiliation{\aff{1} Department of Astrophysical Sciences, Princeton University, Princeton, NJ, 08543
\aff{2} Princeton Plasma Physics Laboratory, Princeton, NJ, 08540}

\begin{document}

\maketitle

\begin{abstract}
Quasisymmetric stellarators are an attractive class of optimised magnetic confinement configurations. The property of quasisymmetry (QS) is in practice limited to be approximate, and thus the construction requires measures that quantify the deviation from the exact property. In this paper we study three measure candidates used in the literature, placing the focus on their origin and the comparison of their forms. The analysis shows clearly the lack of universality in these measures. As these metrics do not directly correspond to any physical property (except when exactly QS), optimisation should employ additional physical metrics for guidance. Especially close to QS minima, treating QS metrics through inequality constraints so that additional physics metrics dominate optimisation is suggested. The impact of different quasisymmetric measures on optimisation is presented through an example.
\end{abstract}

\section{\label{sec:intro} Introduction}
The stellarator concept (\cite{spitzer1958}) has seen a revival in the last decades as an option to achieve controlled thermonuclear fusion. Stellarators are three dimensional magnetic field configurations which confine hot plasmas within. The three-dimensionality of these devices provides the necessary freedom to avoid some of the limiting features that axisymmetric tokamaks suffer from: most notably, current-driven instabilities. However, this freedom can come at a price: the three-dimensionality of the magnetic field might lead to a loss of confinement that continuous symmetry confers upon the tokamak. \par
Quasisymmetry (QS) (\cite{boozer1983,nuhren1988,tessarotto1996,Helander2014,rodriguez2020,burby2020}) is a property of the field that attempts to preserve in 3D stellarators the good neoclassical properties of the tokamak. The promise of this property has led to a search of designs that bear it. Although theoretical work suggests that in an ideal, static equilibrium with isotropic pressure one may not construct exact QS solutions (\cite{garbooz1991ne,landreman2018a,rodriguez2020i}), one may design such configurations approximately. In practice there have been two main approaches to finding approximately quasisymmetric equilibria. One is to directly construct approximate solutions by employing an asymptotic expansion near the magnetic axis (\cite{garbooz1991ne,landreman2018a,landreman2019,rodriguez2020ii}) or near axisymmetry (\cite{plunk2018}). The other, and most commonly employed approach, is to treat the construction of QS configurations as part of an optimisation problem (\cite{bader2019,henneberg2019}). \par
To proceed with optimisation it is necessary to construct cost functions that penalise deviations from QS. This has been done in a number of contexts and forms (\cite{bader2019,henneberg2019,paul2020}). However, at the present time there appears to be a lack of comparative studies of the different forms of cost functions used. The purpose of this paper is to re-evaluate carefully the origin of some of these commonly-used forms and investigate their differences, significance and content. \par
The paper is organised as follows. In Section 2 three different measures of QS are considered. From these basics, the most apparent formal aspects of the metrics are discussed. In Section 3 the measures are formally compared on the same footing. The connection between physical properties associated with QS and the form of the metrics is then explored. Section IV presents a more practical perspective on the measures by looking at the universality of QS measures (how they compare across different near-QS designs) and the importance of the cost functions for optimisation in practice. We close the paper with some concluding remarks and suggestions for future optimisation.

\section{\label{sec:QSintro} Measures of quasisymmetry}
Let us start by defining the concept of \textit{quasisymmetry} from the fundamental perspective of single-particle dynamics. The class of magnetic fields that grants an approximate conserved momentum to the gyrocentre dynamics of charged particles is called \textit{quasisymmetric}. In particular, we focus our attention on \textit{weakly quasisymmetric} magnetic fields as recently defined (\cite{rodriguez2020,constantin2021}). Formally, a magnetic field $\mathbf{B}$ is weakly quasisymmetric iff there exists a symmetry vector field $\mathbf{u}$ such that (\cite{rodriguez2020}),
\begin{gather}
    \mathbf{u}\cdot\nabla B=0, \label{eqn:C1}\\
    \mathbf{B}\times\mathbf{u}=\nabla\Phi, \label{eqn:C2}\\
    \nabla\cdot\mathbf{u}=0, \label{eqn:C3}
\end{gather}
where $\Phi$ labels magnetic flux surfaces that we take to be nested (\cite{rodriguez2021a}). We are assuming the electrostatic potential in the problem to be symmetric. \par
Although this definition has the benefit of directly emerging from single-particle considerations, it is difficult to use it practically to assess whether a field is quasisymmetric or not. Doing so would require solving for the field $\mathbf{u}$. Instead, we would like to have a condition that only involves magnetic field quantities directly. This can be achieved in a straightforward way by constructing $\mathbf{u}$ such that it satisfies Eqs.~(\ref{eqn:C1}) and (\ref{eqn:C2}), $\mathbf{u}=\nabla\Phi\times\nabla B/(\mathbf{B}\cdot\nabla B)$. At the singular points where $\mathbf{B}\cdot\nabla B= 0$ this construction breaks down unless $\nabla\Phi\times\nabla B\cdot\mathbf{B}=0$. (Note that this follows from the conditions (\ref{eqn:C1}) and (\ref{eqn:C2}).) The vanishing of $\nabla\Phi\times\nabla B\cdot\mathbf{B}$ when $\mathbf{B}\cdot\nabla B= 0$ is referred to as \textit{pseudo-symmetry} (\cite{mikhailov2002,skovoroda2005}), and it guarantees that all of the contours of constant $|\mathbf{B}|$ on a flux surface are linked to the torus in the same way, avoiding local extrema on the surfaces (and thus providing well-behaved streamlines of $\mathbf{u}$). Breaking pseudo-symmetry leads to a diverging orbit width (\cite{rodriguez2020}) (i.e., loss of deeply-trapped and barely-trapped particles) at these $|\mathbf{B}|$ extrema. In a QS solution, the construction of $\mathbf{u}$ is then well defined because QS implies PS.
 \par
With the given construction of $\mathbf{u}$, we may substitute it into Eq.~(\ref{eqn:C3}) to obtain
\begin{equation*}
    \frac{\nabla\Phi\times\nabla B\cdot\nabla(\mathbf{B}\cdot\nabla B)}{(\mathbf{B}\cdot\nabla B)^2}=0.
\end{equation*}
This yields the \textit{triple vector formulation} of QS. Namely, if we define the scalar quantity,
\begin{equation}
    f_T=\nabla\psi\times\nabla B\cdot\nabla(\mathbf{B}\cdot\nabla B), \label{eqn:fT}
\end{equation}
the magnetic field is quasisymmetric iff $f_T=0$. (By continuity, this includes pseudo-symmetry.) \par
For the purpose of this paper, all magnetic configurations have well-behaved nested flux surfaces and are non-isodynamic. By isodynamic we mean a field whose magnitude $|\mathbf{B}|$ is a flux function (\cite{bernardin1986}). With these provisions, $f_T$ can be evaluated for non-quasisymmetric configurations and serves as a measure of QS. We have here used $\psi$, the toroidal magnetic flux, as a label of flux surfaces rather than the arbitrary flux function $\Phi(\psi)$. This is one of many possible choices, and one could, for instance, replace $\nabla \psi$ with  $\hat{\mathbf{n}}$ (the normal unit vector to the flux surface) in evaluating \eqref{eqn:fT} on a magnetic surface within a region of continuously nested surfaces.  \par
The measure $f_T$ is a coordinate-independent, scalar measure of the departure from QS. It is a local quantity, although it also has some global properties. For instance, $\langle f_T\rangle=0$, where $\langle \dots \rangle$ is the flux-surface average. \par
We have seen how (\ref{eqn:fT}) results from the fundamental definition of QS in Eqs.~(\ref{eqn:C1})-(\ref{eqn:C3}). This grants $f_T$ the special feature that its application is not tied to a particular form of equilibria. This could be relevant for constructing cost-functions in so-called one-shot optimisation schemes (\cite{Akccelik2006,Dekeyser2014}), i.e., a formulation in which finding equilibria is part of the optimisation problem. Although QS may be explored in the presence of forces more general than isotropic plasma pressure, $p$, it is standard to adopt the \textit{magnetohydrostatic} (MHS) assumption $\mathbf{j}\times\mathbf{B}=\nabla p$ where $\mathbf{j} = \nabla \times \mathbf{B}$ is the current density. Following standard practice, we shall do so for the remainder of this work. \par
A key consequence of this form of equilibria is that $\mathbf{j}\cdot\nabla\psi=0$. Thus, there exist Boozer coordinates (\cite{boozer1981,Helander2014}) $\{\psi,\theta,\phi\}$. In these straight-field line coordinates the covariant form of the magnetic field is $\mathbf{B}=I(\psi)\nabla\theta+G(\psi)\nabla\phi+B_\psi\nabla\psi$ and the coordinate Jacobian $\mathcal{J}=(G+\iota I)/B^2$ where $\iota$ is the rotational transform. Equation~(\ref{eqn:fT}) may then be written explicitly in the form,
\begin{equation*}
    f_T=
   \left[\partial_\theta B\partial_\phi-\partial_\phi B\partial_\theta\right](\partial_\phi+\iota\partial_\theta)B.
\end{equation*}
The QS condition can be rewritten in the form $(\partial_\theta B) (\partial_\phi+\iota\partial_\theta)(\partial_\phi B/\partial_\theta B)=0$, from which it follows that $\partial_\phi B/\partial_\theta B$ must be a flux function. This implies $|\mathbf{B}|$ has an explicit symmetry and the conventional definition of QS then follows (\cite{rodrigGBC,boozer1983}).  We refer to it as the Boozer formulation of QS: 
a magnetic field is QS iff the magnetic field magnitude can be written as a function $B=B(\psi,M\theta-N\phi)$, where $N,M\in \mathbb{Z}$. It is convenient to define the symmetry helicity $\tilde{\alpha}=N/M$ and a helical angle $\chi=\theta-\tilde{\alpha}\phi$. (We will generally consider all forms of symmetry other than quasi-poloidal symmetry.) 
There is no unique way to construct a scalar measure that quantifies how close a given magnetic field is to QS in the presented formulation. It is however customary to define the minimal measure,
\begin{equation}
    f_B={\sum_{\substack{n,m \\ n\neq\tilde{\alpha}m}}}|B_{nm}|^2. \label{eqn:fB}
\end{equation}
Here $B_{nm}$ stands for the Fourier components of the magnetic field magnitude $B$ in Boozer coordinates, and the sum is over the non-symmetric components. All the mode components are weighted equally in this form (although this could be modified). In practice, only the lower-order modes significantly contribute to $f_B$, a result that follows from the smoothness of $|\mathbf{B}|$: an $s$ times differentiable function in $L^1$ on a torus will have Fourier coefficients that generally scale like $\hat{f}(m)\sim1/|m|^s$ (\cite{grafakos2008}). \par
Unlike $f_T$, the helicity of the symmetry $\tilde{\alpha}$ is required for the evaluation of $f_B$. Such knowledge is needed to avoid the symmetric modes of $|\mathbf{B}|$ in the summation of (\ref{eqn:fB}). In a sense, therefore, $f_B$ measures deviations from a particular form of QS. The possibility to enforce a particular helicity can be of practical importance when optimising for QS configurations. \par
This measure is also different from $f_T$ in other two important respects. First, $f_B$ is a single flux surface scalar instead of a local measure. Second, as the Fourier mode resolution of $|\mathbf{B}|$ needed for Eq.~(\ref{eqn:fB}) is in Boozer coordinates, an evaluation of $f_B$ requires explicit knowledge of these coordinates. This prevents application of $f_B$ to scenarios requiring $\mathbf{j}\cdot\nabla\psi\neq0$, unlike $f_T$. The need to compute Boozer coordinates and Fourier resolve $B$ also imposes an additional numerical burden (e.g., using existing codes such as BOOZXFORM) (\cite{sanchez2000,paul2020}). \par
While $f_T$ does not require a Boozer coordinate transformation, it also does not allow specification of the desired helicity, $\tilde{\alpha}$. It is thus of interest to explore alternative measures of QS that incorporate both aspects. We shall now obtain what we call the \textit{two-term formulation} of QS (\cite{paul2020,Helander2014}). \par
We begin from Eqs.~(\ref{eqn:C1})-(\ref{eqn:C3}) again. Defining the inner product $C=\mathbf{B}\cdot\mathbf{u}/\Phi'$ and using Eqs.~(\ref{eqn:C1}) and (\ref{eqn:C2}), we can write $C\mathbf{B}\cdot\nabla B= \mathbf{B}\cdot\nabla \psi\times\nabla B$. This expression holds even when $\mathbf{B}\cdot\nabla B=0$. To involve $C$ directly in the remaining Eq.~(\ref{eqn:C3}), we rewrite it in the form (\cite{rodrigGBC}),
\begin{equation}
    \mathbf{B}\cdot\nabla(\mathbf{B}\cdot\mathbf{u})=\mathbf{j}\cdot\nabla\Phi.
\end{equation}
Under the assumption of $\mathbf{j}\cdot\nabla\psi=0$, it follows that $C=C(\psi)$ must be a flux function. We can then write
\begin{equation}
    f_C=\mathbf{B}\cdot\nabla \psi\times \nabla B-C(\psi) \mathbf{B}\cdot\nabla B, \label{eqn:fC}
\end{equation}
which must vanish for some flux function $C$ for a QS field. This is the \textit{two-term formulation} of QS. To avoid having to perform a search for $C$ every time (\ref{eqn:fC}) is evaluated, we need some form for the flux function $C$ in terms of more recognisable quantities. To do so, we consider the limit of a QS configuration and adopt Boozer coordinates. In that case, and for $\partial_\chi B\neq 0$ (otherwise $f_C=0$ for any value of $C$),
\begin{equation}
    C=\frac{\mathbf{B}\cdot\nabla \psi\times\nabla B}{\mathbf{B}\cdot\nabla B}=\frac{G+\tilde{\alpha} I}{\iota-\tilde{\alpha}}. \label{eqn:Ceq}
\end{equation}
(This flux function $C$ is often given the symbol $F$ in the literature (\cite{Helander2014,paul2020})). Now $C$ is expressed in terms of physical quantities: the Boozer currents $I$ and $G$ and the rotational transform $\iota$. The currents may be calculated without resorting to Boozer coordinates directly (see Eq.~(16) in (\cite{Helander2014})). Any flux coordinate system will do, under the assumption of well-defined, nested flux surfaces. \par
It is clear that the helicity of the symmetry $\tilde{\alpha}$ needs to be specified, just as one must do to evaluate $f_B$. The function $C$ for which it is needed has important physical meaning (\cite{rodriguez2020}), as it is a direct measure of the banana width of trapped particles, $\Lambda$, in the QS limit. In more detail, $\Lambda|\nabla\psi|\sim \rho_\parallel C$, where $\rho_\parallel$ is the Larmor radius associated with the parallel velocity. The resonance at the surface $\iota=\tilde{\alpha}$ leads to $C\rightarrow \infty$, unless $\mathbf{B}\cdot\nabla\psi\times\nabla B=0$ on the surface. This latter condition forces the surface to be isodynamic,
which also prevents the appearance of a current singularity and the potential opening of a magnetic island (\cite{rodriguez2021a}). In the present work we shall avoid such surfaces altogether. \par
In summary, we have a measure $f_C$ that is local like $f_T$, for which the helicity of the symmetry must be prescribed as in $f_B$, and for which no Boozer coordinates are needed. This formulation is also more amenable to gradient-based optimisation techniques such as adjoint methods (\cite{paul2020}). \par
This constitutes the derivation of the three measures of QS—$f_B$, $f_T$ and $f_C$—that will occupy us in this paper. In Table \ref{tab:QSmeas} we summarise the main formal and construction differences of the three forms. \par
\begin{table}
    \centering
    \begin{tabular}{c||c|c|c|c|}
    & \textsc{Helicity} & \textsc{Boozer} & $\mathbf{j}\cdot \nabla\psi=0$ & \textsc{Info} \\\hline\hline
    $f_B$ & I & Y & Y & G \\
    $f_C$ & I & N & Y & L \\
    $f_T$ & O & N & N & L \\
    \end{tabular}
    \caption{\textbf{Basic properties of QS measures.} Table summarising the basic formal features of the QS measures introduced in Section 2. These include: whether the helicity needs to be specified as an input (I) or not (O), whether Boozer coordinates are required to evaluate it, whether $\mathbf{j}\cdot\nabla\psi=0$ is a necessary assumption and whether the measure is local (L) or global (G).}
    \label{tab:QSmeas}
\end{table}

\section{\label{sec:QScomparison} Comparison of cost functions}
All three forms introduced in the previous section can be interpreted as measures of QS: they all share the basic feature that they vanish only for a configuration that is exactly quasisymmetric. However, each form treats deviations from QS differently. The focus of this section is to evaluate how these measures compare with each other formally, and to connect them to some of the physical properties usually associated with QS.

\subsection{Formal relation}
It is convenient
to measure deviations from QS in terms of the asymmetric components of the magnetic field magnitude. Thus, to understand the forms of $f_T$ and $f_C$, it is informative to learn how the different Fourier modes of $|\mathbf{B}|$ are weighted in each of these metrics. Then a comparison to $f_B$ may be done on an equal footing. \par
Let us start with the definition of $f_C$ and write the magnitude of the magnetic field as a Fourier series,
\begin{equation*}
    B=\sum_{n,m}B_{nm}e^{i(m\theta-n\phi)},
\end{equation*}
where the angular Boozer coordinates $\theta$ and $\phi$ are used for the Fourier resolution of the field. Using the construction of $C$ in Eq.~(\ref{eqn:Ceq}) we may write after some algebra,
\begin{equation}
\begin{aligned}
    f_C=&\frac{B^2}{\tilde{\alpha}-\iota}(\partial_\phi+\tilde{\alpha}\partial_\theta)B\\
    =&i\frac{B^2}{\iota-\tilde{\alpha}}\sum_{n/m\neq\tilde{\alpha}}(n-m\tilde{\alpha})B_{nm}e^{i(m\theta-n\phi)}. \label{eqn:compfEfFloc} 
\end{aligned}
\end{equation}
The first line shows that $f_C/B^2$, although a local measure, has some global properties like $f_T$ does. In particular, the integral along closed streamlines of $\mathbf{Q}=\mathcal{J}\nabla\psi\times\nabla\chi$ must vanish, that is, $\oint (f_C/B^2)\mathrm{d}l/|\mathbf{Q}|=0$. 
The function $f_C/B^2$ is also clearly linear in the Fourier content of $B$, but interestingly, not all modes are weighted equally. The modes `furthest' from the symmetry direction are weighted more heavily than those close to it. 
\par
Although we have successfully written $f_C$ in terms of the Fourier content of $B$, its form is still far from that of $f_B$. To make a closer comparison to $f_B$ we need to construct a single flux-surface scalar from $f_C$. We take Eq.~(\ref{eqn:compfEfFloc}) and write it as,
\begin{equation*}
    B_{nm}=-i\frac{\iota-\tilde{\alpha}}{n-m\tilde{\alpha}}\left(\frac{f_C}{B^2}\right)_{nm},
\end{equation*}
where $(\dots)_{nm}$ represents the $(n,m)$ Fourier component of the function in brackets in Boozer coordinates. Using Parseval's theorem, we write
\begin{equation*}
    \sum_{n,m}(n-m\tilde{\alpha})^2|B_{nm}|^2=\frac{(\iota-\tilde{\alpha})^2}{(2\pi)^2}\int\mathrm{d}\theta\mathrm{d}\phi\left(\frac{f_C}{B^2}\right)^2,
\end{equation*}
which can be rewritten as,
\begin{equation}
    \overline{\left(\frac{f_C}{B^2}\right)^2}=\sum_{n,m}\left(\frac{n-m\tilde{\alpha}}{\iota-\tilde{\alpha}}\right)^2|B_{nm}|^2, \label{eqn:relfFfE}
\end{equation}
where $\overline{(\dots)}$ denotes the $(0,0)$ Fourier component of the expression in brackets. Equation (\ref{eqn:relfFfE}) is now in a form close to $f_B$. The scalarised version of $f_C$, in this case normalised to $B^2$ and averaged over the surface, has a form similar to $f_B$ in that it involves a sum over the non-symmetric Fourier modes of $B$. The main difference comes through the different weighting of the modes. The measure $f_C$ can be thought of as a modification of $f_B$ in which asymmetric components of the magnetic field magnitude furthest from QS are more heavily penalised. \par
The difference in weights implies that $f_B$ and $f_C$ are actually not monotonic with respect to one another. Changing the energy content of the Fourier modes may lead to $f_B$ increasing but $f_C$ decreasing (and vice-versa). Having said that, there also is a class of changes for which both metrics respond in the same way. Given that all terms in the summation of Eqs.~(\ref{eqn:fB}) and (\ref{eqn:relfFfE}) are positive, reducing any individual mode $|B_{nm}|$ will, for instance, lead to the reduction of both metrics. This observation has important implications for the universality of QS measures and optimisation, to be explored in the next sections. \par
Let us now see how $f_T$ fares in comparison. When expressed in terms of the Fourier content of the magnetic field strength, the non-linear character of the triple product formulation is clear. We explicitly write,
\begin{equation}
    (\mathcal{J}^2f_T)_{nm}=i\sum_{k,l}(\iota l-k)(km-nl)B_{n-k,m-l}B_{k,l},
\end{equation}
where $\mathcal{J}$ is the Jacobian of Boozer coordinates.
The non-linearity appears in the form of a convolution that mixes modes in a non-trivial way. We also have one higher derivative of the field strength compared to $f_B$ or $f_C$. Scalarising this expression by computing an average like in the case of $f_C$ does not help in bringing this form any closer to $f_B$. The convolution couples modes together, making it unclear whether reducing a single asymmetric mode will reduce $f_T$ as it did for $f_C$ and $f_B$. \par
We may try to gain some additional understanding on how $f_T$ is affected by changes in the mode content by considering a system that lies close to QS. Consider a small deviation $\tilde{B}$ in the field magnitude from a dominantly QS field. Then a linearisation of $f_T$ takes the form,
\begin{equation*}
    (\mathcal{J}^2f_T)_{nm}=i(\tilde{\alpha}m-n)(\iota m-n)\sum_llB_{\tilde{\alpha}l,l}\tilde{B}_{n-\tilde{\alpha}l,m-l}.
\end{equation*}
It is clear from this expression that decreasing $\tilde{B}$ decreases the magnitude of $f_T$. However, even upon linearisation it is not obvious that reducing a particular Fourier harmonic will guarantee a lower value of $f_T$. \par
The measure $f_T$ is, in this form, distinct from the previous metrics. However, there is actually a natural relation between the triple vector form and $f_C$ through a simple magnetic differential equation (see Appendix A). This is reasonable given the common origin of the metrics.

\subsection{Physical relations}

We have explored three forms for QS in this work. These constructions describe deviations from QS differently. To understand the potential physical meaning of these differences, we now explore physical phenomena often associated with QS configurations and their relation to the cost functions. \par

\paragraph{Single-particle dynamics}
The concept of QS as presented in this work is built on the single-particle dynamics. In Section 2, beginning with the most fundamental definition, we constructed three measures of QS. Thus, it is natural to expect that all these metrics should have some relation to the single-particle dynamics. \par
Let us start from the definition of the momentum $\Bar{p}=-\Phi+v_\parallel\mathbf{b}\cdot\mathbf{u}$, where $v_\parallel$ is the parallel velocity of the particle under consideration (\cite{rodriguez2020}). For a QS configuration where $\mathbf{u}$ satisfies Eqs.~(\ref{eqn:C1})-(\ref{eqn:C3}), this momentum is approximately conserved to order $O(\rho/L)$, where $\rho$ is the particle Larmor radius and $L$ a characteristic length of the system. When the symmetry is broken, we may ask how this momentum evolves in time. To describe that evolution we need to define the vector field $\mathbf{u}$. We can always choose this field in a way that Eqs.~(\ref{eqn:C1}) and (\ref{eqn:C2}) are satisfied by construction. To do so we shall assume the existence of flux surfaces as well as pseudo-symmetry (\cite{mikhailov2002,skovoroda2005}). The latter guarantees that the contours of constant $|\mathbf{B}|$ are linked to the torus avoiding local extrema on the surfaces. These well-behaved contours are vital to the notion of a continuous symmetry, and necessary to prevent $\mathbf{b}\cdot\mathbf{u}\rightarrow\infty$. Once $\mathbf{u}$ has been constructed, one evaluates the dynamics of $\bar{p}$ using the leading order form of the equations of motion from the Littlejohn Lagrangian (\cite{rodriguez2020,littlejohn1983}). We may thus write to $O(\rho/L)$,
\begin{equation}
    \frac{\mathrm{d}}{\mathrm{d}t}\Bar{p}=\frac{v_\parallel^2}{B^2}\left[\mathbf{B}\cdot\nabla(\mathbf{B}\cdot\mathbf{u})-\mathbf{j}\cdot\nabla\psi\right]=-\left(\frac{v_\parallel}{\mathbf{B}\cdot\nabla B}\right)^2f_T. \label{eqn:dpdtfT}
\end{equation}
This shows that $f_T$ constitutes a measure of the conservation of $\bar{p}$. Of course, even if $f_T\neq0$ the above could result in conservation of $\bar{p}$ in the bounce-averaged sense. This averaged conservation could be seen as the difference between QS and \textit{omnigeneity} (\cite{landreman2012,Helander2014}). It is also important to note from (\ref{eqn:dpdtfT}) that the cost function $f_T$ appears with a factor of $1/(\mathbf{B}\cdot\nabla B)^2$. This factor penalises deviations from QS close to the extrema of $|\mathbf{B}|$ along field lines. The special importance of the regions close to the extrema is reminiscent of the requirements for the confinement of barely and deeply trapped particles in omnigeneity. The presence of this resonance makes the combination $f_T/(\mathbf{B}\cdot\nabla B)^2$ numerically ill-behaved as a modified cost function. A practical implementation could perhaps be achieved by regularizing the combination, as has been employed in optimisation for pseudo-symmetry (\cite{mikhailov2002}). Although attractive from the single-particle perspective, and even if the resonance is amended, the measure would still exhibit sharp gradients due to the heavy weight near extrema of $|\mathbf{B}|$ along field lines.  \par
\begin{figure}
    \centering
    \includegraphics[width=0.45\textwidth]{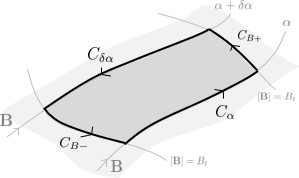}
    \caption{\textbf{Schematic of the loop integral for $\partial_\alpha \mathcal{J}_\parallel$.} Schematic of the path integral followed to compute the field line dependence of the second adiabatic invariant $\mathcal{J}_\parallel$. The contours match field lines and contours of constant $|\mathbf{B}|=B_t$.}
    \label{fig:diagdalphJpar}
\end{figure}

The argument related to the dynamics of $\bar{p}$ might appear to some extent artificial, as $\bar{p}$ really only gains special dynamical meaning in the limit of QS. 
We may alternatively look at a more physically relevant property: the net drift of magnetised particles off flux surfaces. It is convenient to express the off-surface bounce-averaged drift of particles, $\Delta\psi$, as $\partial_\alpha \mathcal{J}_\parallel$, where $\mathcal{J}_\parallel=\oint v_\parallel\mathrm{d}l$ is the second adiabatic invariant (calculated taking the integral along the field line and between bouncing points) and $\alpha$ is the usual magnetic field-line label (\cite{Helander2014}). To evaluate this derivative of $\mathcal{J}_\parallel$, consider two nearby lying magnetic field lines on the same magnetic surface at $\alpha$ and $\alpha+\delta\alpha$ (see Fig.~\ref{fig:diagdalphJpar}). Take the turning points that define the integral for $\mathcal{J}_\parallel$ to be at a field magnitude $|\mathbf{B}|=B_t$. We shall assume that these bounce points exist on both of these field lines (as we are looking at the $\mathcal{J}_\parallel$ associated to a class of particles defined by its bounce points). Then taking the line integral along the contour in Fig.~\ref{fig:diagdalphJpar} we compute,
\begin{equation*}
    \left(\int_{C_\alpha}+\int_{C_{\delta\alpha}}+\int_{C_{B+}}+\int_{C_{B-}}\right)v_\parallel\mathbf{b}\cdot\mathrm{d}\mathbf{l}=\delta\alpha\int \nabla\psi\cdot\nabla\times(v_\parallel \mathbf{b})\frac{\mathrm{d}l}{B}.
\end{equation*}
To obtain the equality, we apply Stokes' theorem and take the limit of small $\delta\alpha$, taking the latter integral with respect to length along a field line, $l$. Assuming that $\mathbf{j}\cdot\nabla\psi=0$ and noting that the integrals along $C_{B+}$ and $C_{B-}$ vanish as $v_\parallel=0$ wherever $|\mathbf{B}|=B_t$, we may rewrite,
\begin{align*}
    \partial_\alpha \mathcal{J}_\parallel=-\int\frac{\nabla\psi\times\nabla B\cdot\mathbf{B}}{\mathbf{B}\cdot\nabla B}\frac{1}{B^2}\left(\frac{E}{v_\parallel}+\frac{v_\parallel}{2}\right)\mathrm{d}B,
\end{align*}
where $B$ is being used to parametrise the integral along the field line. To be precise, we should split the integral along the field line into concatenated pieces joint at $|\mathbf{B}|$ extrema. The vanishing of $\mathbf{B}\cdot\nabla B$ at the stitching points (and the consequent divergence of the integrand) is generally not problematic as one can consider switching back to integrating with respect to $l$ arbitrarily close to them, with that small piece of the integral constituting a vanishingly small contribution. We shall drop this distinction for simplicity. \par
One may write the kinematic expression $(E/v_\parallel+v_\parallel/2)/B^2=\mathbf{B}\cdot\nabla(v_\parallel/B)/(\mathbf{B}\cdot\nabla B)$. As a result, any flux function multiplying this kinematic expression will lead to a vanishing integral. Thus, the first fraction in the integrand can be written as $f_C/(\mathbf{B}\cdot\nabla B)$. This shows that $f_C$ has a direct relation to the field-line dependence of the second adiabatic invariant. The complicated nature of the kinematic factor however obscures this relation. Proceeding with integration by parts and using Eq.~(\ref{eqn:fT2fP}), 
\begin{align}
    \partial_\alpha \mathcal{J}_\parallel=\left.\frac{\nabla\psi\times\nabla B\cdot\mathbf{B}}{\mathbf{B}\cdot\nabla B}\frac{v_\parallel}{B}\right|_{v_\parallel=0}+\int\frac{f_T}{(\mathbf{B}\cdot\nabla B)^2}v_\parallel\mathrm{d}l. \label{eqn:daJ}
\end{align}
Thus, we have constructed a simple relation between the bounce-averaged, off-surface drift $\Delta\psi$ and the QS measure $f_T$. The measure comes once again normalised by a factor $(\mathbf{B}\cdot\nabla B)^2$. However, this $f_T$ term is not the only term in (\ref{eqn:daJ}). 
For the expression in (\ref{eqn:daJ}) to be finite (or vanish as we want for QS), the boundary term in (\ref{eqn:daJ}) should not diverge, which requires $\nabla\psi\times\nabla B\cdot\mathbf{B}=0$ wherever $\mathbf{B}\cdot\nabla B=0$. This condition is precisely the condition of pseudo-symmetry introduced earlier (\cite{mikhailov2002,skovoroda2005}). Note that this pseudo-symmetry condition appears separate from the $f_T$ integral in (\ref{eqn:daJ}). Thus, generally, even when $f_T\neq0$ as in ominigeneous configurations, the condition should be satisfied if $\partial_\alpha\mathcal{J}_\parallel$ vanishes (and deeply trapped and barely trapped particles are confined). As discussed in Section 2, requiring $f_T=0$ everywhere includes this condition. Guaranteeing the boundary term vanishes also ensures the existence of bouncing points $B_t$ on nearby field lines, in agreement with the assumptions made in evaluating $\partial_\alpha \mathcal{J}_\parallel$. \par
The discussion above shows that both $f_T$ and $f_C$
can be regarded as a natural measure of QS in the single-particle perspective. However, when the configuration fails to be pseudo-symmetric, the interpretation of $f_T$ is obscured. \par
The measure $f_C$ offers some additional particle dynamics insight. As we mentioned when constructing $f_C$ in Sec.~2, the function $C$ in $f_C$ is a measure of the banana width of bouncing particles in the QS limit. The expression in (\ref{eqn:compfEfFloc}) inherits the scaling $\propto 1/\bar{\iota}$ where $\bar{\iota}=\iota-\tilde{\alpha}$ from $C$.
Hence, $f_C$ also has scaling related to the orbit width and the bootstrap current, which is proportional to the banana width (\citep{Helander2014}). \par
Although the metrics bear some important physical content in their relation to the behaviour of single-particle dynamics, the mapping is not one-to-one. More sophisticated measures are required for an accurate description of the dynamics of single particles. Especially important are alpha particles, for which complicated metrics (see (\cite{bader2019})) or expensive simulations are needed.

\paragraph{Neoclassical transport} 
Let us now consider the important physics of \textit{neoclassical transport} (also related to enhanced neoclassical viscosity). One of the distinguishing features of QS is to grant a low level of neoclassical transport to the magnetic configuration (\cite{boozer1983}). In particular, it prevents the detrimental $1/\nu$ transport that tends to spoil confinement in stellarators. Of course, as we deviate from QS, this property will degrade. We would like to understand how this degradation is captured (if at all) by the different metrics of QS presented in the paper. \par
To study this, we shall introduce first the classical stellarator concept of helical ripple. We define helical ripple as $\epsilon_h\sim B_{nm}/B$, a measure of the symmetry-breaking magnetic field magnitude. This quantity is clearly linked to $f_B$ in (\ref{eqn:fB}). When a single, small asymmetric mode on top of a QS field is considered, in the $1/\nu$ and $\sqrt{\nu}$ regimes, this ripple leads to an enhancement of neoclassical losses that scales roughly as $D\sim\epsilon_h^{3/2}$ (\cite{ho1987,nemov1999,helander2008}). Therefore, there is a direct relationship between $\epsilon_h$ and transport. In this single-mode classical stellarator scenario $f_B$ \textit{is} the ripple (unlike in the more general case). It is then clear that $f_B$ does incorporate some direct information about transport.  \par
This classical view on transport has the drawback of treating all `ripples' on the same footing. However, a more detailed approach shows that different modes contribute distinctly (\cite{Calvo2014}). 
Transport in the $1/\nu$ regime is driven by bouncing particles near minima of $|\mathbf{B}|$ along field lines. The enhancement of transport can then come from the introduction of significant additional wells in which new bouncing particles will live. 
However, this is not the only mechanism that can enhance transport. As particles precess following constant $\mathcal{J}_\parallel$ surfaces, a significant difference in the $|\mathbf{B}|$ profile between field lines on the same flux surface will lead to an enhancement in the drifting of bouncing particles off the surface. (Additional effects such as collisional detrapping may also occur.) These phenomena are quantified in (\cite{Calvo2014}). Small deviations from QS with helical ripple $\epsilon_h$ that do not affect the $|\mathbf{B}|$ profile significantly lead to a transport enhancement $\langle\mathbf{\Gamma}\cdot\nabla\psi\rangle\sim\epsilon_h^2$. (The same scaling is anticipated for flow damping and generally deviations from intrinsic ambipolarity.) Now, if the deviation is due to a mode with a large gradient along $\mathbf{B}$, then the changes lead to an enhancement by a factor $O(\epsilon_h)$. When the deviations lead to significant differences between field lines, then the transport changes significantly, not amenable to a perturbative treatment. (Aspects of this scenario can be understood from an analysis of the second adiabatic invariant, studied in the previous section.) \par
In this more detailed picture it is clear that different modes of $|\mathbf{B}|$ contribute differently to transport. The enhancement is significant when the gradients along the field line are large in the sense $|\mathbf{b}_0\cdot\nabla B_1|\sim|\mathbf{b}_0\cdot\nabla B_0|$, with $B_1$ the asymmetric field strength. In a more illuminating form,  
$|\mathbf{b}_0\cdot\nabla B_1|/|\mathbf{b}_0\cdot\nabla B_0|\sim \left[(n-\iota m)/\left(M(\tilde{\alpha}-\iota)\right)\right](\epsilon_h /\epsilon_\mathrm{QS})$ where we have taken $\epsilon_{\text{QS}}$ to represent the helical ripple associated to the symmetric part of $B$ with symmetric poloidal mode number $M$. The weight in the numerator indicates the predominance of modes with large gradients along the field line, while the denominator serves as a comparative measure for the the gradient along the field line due to the symmetric field. \par
This difference between modes is clearly missed by the construction of $f_B$. However, we have shown that the angle-averaged $(f_C/B^2)^2$ provides a weighted version of $f_B$ in which higher-order modes are increasingly penalised $\propto(n-m\tilde{\alpha})^2$. Although this unequal weight provides a distinction between modes, the weight is not of the form $(n-\iota m)$ as in the transport criterion. It instead weights higher-order modes according to their similarity with respect to the symmetry direction. Thus, $f_C$ only qualitatively matches the high-order mode weighting of the transport scaling. The weight that appears in the definition of $f_C$ can be interpreted as a consequence of the smoothness of the function. The quasisymmetric modes are excluded from the sum through a weight that can be written as a differential operator: the weight must be zero at $n/m=\tilde{\alpha}$ and smoothly increasing away from it. It thus appears natural that the weight in $f_C$ is centred about the symmetry direction. This requirement is avoided in $f_B$, where the symmetric discrete modes are avoided. To more accurately treat the impact on transport, one could imagine a modified version of $f_B$ in which the mode weights are given by $(n-\iota m)$ and the symmetric components are excluded from the summation. It would, however, still require Boozer coordinates (unlike $f_C$). Interestingly, the weight $(m\iota - n)$ appears in the linearised version of $f_T$ in Sec.~3.1. \par
So far we have focused on small, isolated deviations from QS. In general, however, transport calculations are significantly more complex. The presence of multiple ripples will lead to intricate changes in the trapping and passing particle behaviour. There exist more sophisticated metrics that attempt to describe these changes in more detail. A commonly used example is the $1/\nu$-transport regime effective ripple, $\epsilon_\mathrm{eff}^{3/2}$ (\cite{nemov1999,nemov2014}). The complexity of this measure is not reflected in either $f_B$ or $f_C$. (See the following sections.) Thus, although $f_B$ and $f_C$ do contain relevant information regarding transport, they cannot be taken as accurate measures of it. \par

\paragraph{Current singularities and islands}
Although flux surfaces are assumed throughout our analysis, we may also relate the QS metrics to the potential presence of current singularities and magnetic islands (\cite{tessarotto1995,rodriguez2021a}). Fourier modes of $|\mathbf{B}|$ that resonate with the finite rotational transform can lead to Pfirsch-Schl\"{u}ter (PS) current singularities. These finite-$\beta$ currents are associated with the potential opening of magnetic islands at the location of the singularities (\cite{reiman1984,rodriguez2021a}). The PS current singularity arises due to the periodicity constraint on the magnetic differential equation,
\begin{align}
    \mathbf{B}\cdot\nabla\left(\frac{j_{||}}{B}+\frac{p'C}{B^2}\right)=-2p' \frac{f_C}{B^3},
\end{align}
for the parallel current, $j_{||}$. The periodicity condition required for solvability is,
\begin{align}
   p' \oint \frac{f_C}{B^4} \, \mathrm{d}l = 0,
\end{align}
where $l$ measures length along a field line. We note that $f_C$ appears explicitly. This condition will be satisfied in QS, thus precluding the existence of PS singularities (except possibly at the conflictive $\iota=\tilde{\alpha}$ surface, see Sec.~2). Note that solvability may still be obtained away from QS if $p' = 0$ or the line integral vanishes.
\par
One may generally show that in MHS, the resonant radial magnetic field due to the singularities is $B_{\rho nm}\propto (1/B^2)_{nm}$ (in simplified geometry), and thus $W_i\propto\sqrt{(1/B^2)_{nm}}$ (\cite{reiman1984,bhattacharjee1995,rodriguez2021a}). A non-QS field will potentially have contributions at many rational surfaces, as driven by the asymmetric modes. In that sense, any sum over the asymmetric modes of $B$, such as $f_B$ or $f_C$, will include some of this information. 
Resonances only occur when the rotational transform matching the asymmetric mode helicity exists within the plasma, and thus a more faithful metric would only account for the \textit{resonant} asymmetric modes. Due to the generally asymmetric geometry of configurations, resonant fields arising from PS currents may appear at other rational surfaces through toroidal coupling (\cite{rodriguez2021a}). 

\section{\label{se:QSinPractice} Quasisymmetry in practice: universality and cost functions}
The discussion so far has explored the origin of the three different measures of QS—$f_B$, $f_C$ and $f_T$—and discussed their formal properties and related physical content. In this section we explore some more practical aspects of these measures. First, we will address the question `Which configuration is more quasisymmetric?' and discuss the universality of QS. At a second stage, we explore the implications of these different formulations for the optimisation of stellarators through numerical examples.

\subsection{Universality of QS}
\label{sec:universality}
The goal of the metrics $f_B$, $f_C$ and $f_T$ is to quantify the deviation from QS. In this sense, all these forms would appear to be equally valid candidates. Given the variety of choice, we would like to understand to what extent it is meaningful to state that a certain field configuration is more quasisymmetric than another as measured by these metrics. \par
As demonstrated in the comparison of the previous section, QS metrics are not generally monotonic with respect to each other. For example, decreasing $f_B$ does not imply decreasing $f_C$. One could, for instance, exchange the energies of a larger low-order mode to a high-order one without changing $f_B$ but modifying $f_C$. The non-linearity of $f_T$ makes the lack of monotonicity somewhat more marked. This lack of monotonicity makes it difficult to determine magnetic configurations which are `closer' to QS, or for that matter, if the question is meaningful. As different measures of QS yield different answers, any comparison must be defined with respect to a particular standard. The notion thus lacks \textit{universality}: the quantification of closeness to QS is in some sense arbitrary, and does not necessarily have a clear physical meaning attached to it.   \par
Let us consider some practical examples. In order to use the metrics of QS quantitatively, we first need to normalise and scalarise them. Only then will we be able to make fair comparisons across different devices. Perhaps unsurprisingly, we will see that this procedure is not unique, emphasising the lack of universality. \par
Let us start with the metric $f_B$, which measures the Fourier content of $B$ in Boozer coordinates. It is customary to normalise it to the total magnetic field energy. So we define,
\begin{equation}
    \hat{f}_B=\sqrt{\frac{1}{\overline{B^2}}f_B}. \label{eqn:normFb}
\end{equation}
Doing so ensures $\hat{f}_B\in[0,1)$. This finite range of values makes $f_B$ convenient, although the requirement $B>0$ excludes the value of 1.
Although this normalisation seems the most natural, it is not unique. For instance, we could decide to normalise $f_B$ to (only) the symmetric part of the field excluding the flux-averaged part, or we could normalise by the flux-surface average of $B^2$, $\langle B^2\rangle$. For the purpose of this discussion, we will take the natural and more conventional form (\ref{eqn:normFb}). \par
The choice of a `natural' normalisation is unclear for $f_C$ and $f_T$. Fundamentally, our normalisation should give a dimensionless measure. This principle is, of course, not specific enough. To constrain the construction of the measure further, we shall try making their Fourier forms look closest to $f_B$. This is a convenient choice. Let us start with $f_C$, which dimensionally scales as $f_C\sim  B^3$. In addition to this dimensional factor, motivated by the form of Eq.~(\ref{eqn:compfEfFloc}), we also choose to normalise by $1/(\tilde{\alpha}-\iota)$ to bring the measure closer to the form of $f_B$. One could argue for adding additional dimensionless factors such as the inverse aspect ratio, $\epsilon$, which characterises the magnitude
of $C\mathbf{B}\cdot\nabla B$ for $\chi = \theta$, but for simplicity we shall not do so here. In view of the normalisation of $f_B$, we normalise and scalarise $f_C$ as follows,
\begin{equation}
    \hat{f}_C=\sqrt{\frac{\langle f_C^2\rangle}{\langle B^2\rangle^{3}}|\tilde{\alpha}-\iota|^2}.
\end{equation}
This leads to a form that is close to but not identical to Eq.~(\ref{eqn:compfEfFloc}). 
\par
The normalisation of $f_T$ is inherently more complex. The triple vector product introduces explicit length scales which must be non-dimentionalised. Once again, we look at the form of $f_T$ in (\ref{eqn:relfFfE}). It is then natural to normalise and scalarise $f_T$ using,
\begin{equation}
    \hat{f}_T=\sqrt{\langle f_T^2\rangle\frac{R^4}{\langle{B^2}\rangle^4}}.
\end{equation}
In obtaining this normalisation we have used the scaling $\mathcal{J}\sim R/B$, where $R$ is the major radius. Once again, one could argue that the choice is, to some extent, arbitrary. It is interesting to note that the combination $f_T/(\mathbf{B}\cdot\nabla B)^2$, recurrent in the description of single-particle dynamics in the previous section is by construction dimensionless. Thus, it appears to be a natural choice. However, we do not consider it here due to the added complication of the resonant denominators and the associated steep gradients. 
\par
Now that we have chosen specific normalisations for the QS metrics, we may compare their values in stellarator equilibria. We consider two scenarios. First, we will focus on the comparison of the metrics in configurations that have been designed to be approximately quasisymmetric. Then we will study how these measures differ when the system has not necessarily been optimised to be quasisymmetric.  \par
\begin{figure*}
    \centering
    \includegraphics[width=0.8\textwidth]{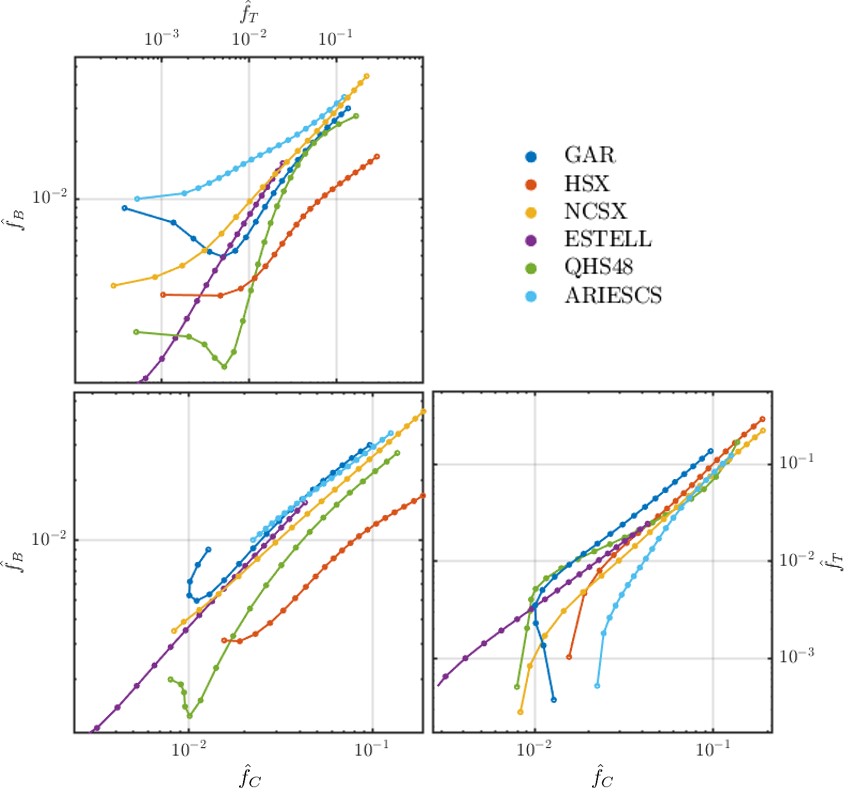}
    \caption{\textbf{Comparison of defined cost functions in quasisymmetric designs.} Plots in logarithmic scale including a comparison of the cost functions $\hat{f}_B$, $\hat{f}_C$ and $\hat{f}_T$ for a number of quasisymmetric devices. The scatter points for each device indicate values at a number of equally spaced magnetic flux surfaces.}
    \label{fig:costFuncComp}
\end{figure*}
We present in Figure \ref{fig:costFuncComp} a comparison of the metrics in six quasisymmetric configurations:\footnote{\label{foot:git_equilibria} https://github.com/landreman/vmec\_equilibria.git} GAR (\cite{Garabedian2008,Garabedian2009}), HSX (\cite{Anderson1995}), NCSX (\cite{Zarnstorff2001}), ESTELL (\cite{Drevlak2013}), QHS48 (\cite{Ku2010}) and ARIESCS (\cite{Najmabadi2008}). It is clear that, within a given device, the correlation between cost functions is quite good. All measures qualitatively show the same behaviour: for each configuration, QS is better realised closer to the axis than away from it. (This trend seems to be preserved except in regions which are very close to the axis.) However, the same cost function correspondence does not seem to hold if we make comparisons between different configurations. \par
\begin{figure}
    \centering
    \includegraphics[width=0.55\textwidth]{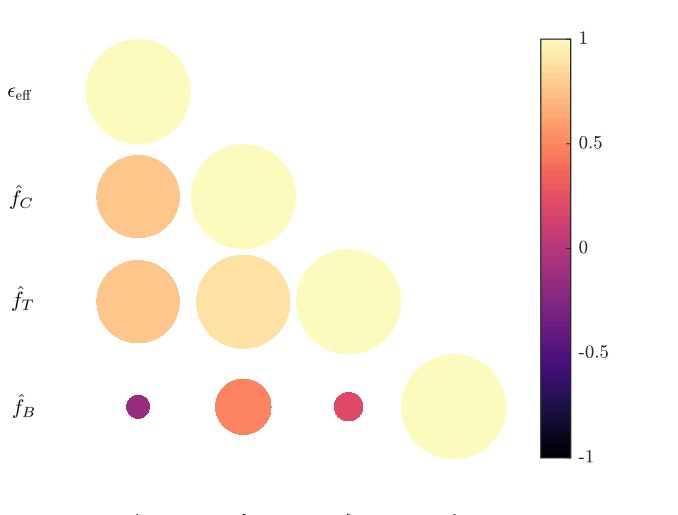}
    \caption{\textbf{Correlation between edge values of cost functions and $\epsilon_\mathrm{eff}$ of different quasisymmetric configurations.} Diagram showing the Spearman correlation between the edge values of the different cost functions and $\epsilon_\mathrm{eff}$ of the configurations in Fig.~\ref{fig:costFuncComp}. The colour represents the coefficient (as does the diameter of the coloured circles). }
    \label{fig:corrCostFuncEps}
\end{figure}

To quantify this lack of ordering across devices, we compute the Spearman correlation coefficient, which considers correlation only taking order into account. We show this in Fig.~\ref{fig:corrCostFuncEps}. The high correlation between cost functions within a particular device (which in all cases was $\sim0.9-1$) clearly does not hold across devices. Although $\hat{f}_C$ and $\hat{f}_T$ retain a high degree of correlation ($\sim 0.9$), the conventional $\hat{f}_B$ measure seems largely uncorrelated to the others. (Of course, in this inter-configuration comparison the normalisation factors chosen are particularly important.) \par

The loss of correlation across machines emphasises the main point on universality from a practical perspective. Depending on the form of cost function taken, one could say either A or B is more quasisymmetric. The statement has only a relative meaning at best for these optimised configurations. Without a physical basis to the measures it is difficult to argue in favour of one definition over the other. For a more meaningful statement, one should instead compare these functions through some physical property. Ultimately we are interested in finding configurations that are close to QS because we seek some of the physical properties that QS confers upon magnetic configurations. In Figure \ref{fig:corrCostFuncEps} we compare $\epsilon_\mathrm{eff}$ (which also contains some arbitrary normalisation factors in its definition (\cite{nemov1999})) to the three cost functions for the devices of Fig.~\ref{fig:costFuncComp}. Other physics measures such as alpha particle confinement could also be suited for such comparison (\cite{bader2019}). There exists some correlation between the transport and the normalised values of $f_C$ and $f_T$. It is especially remarkable the seemingly little correlation between $\hat{f}_B$, which is widely used for optimisation, and $\epsilon_\mathrm{eff}$. Overall, for these `quasisymmetric' configurations, the QS cost functions appear not to be good indicators of their transport levels.
Similar behaviour has been observed in previous studies of quasisymmetric configurations (\cite{Martin2020}).  \par
 \par
In summary, when comparing quasisymmetric configurations, there is no universal measure of QS. The comparison depends on the metric used, and these differ from each another. To make a comparison physically meaningful, especially close to optimised configurations, we are in need of resorting to direct physical measures of the configuration (e.g., neoclassical transport or alpha particle confinement).  \par

\subsection{Optimising for QS}
The differences in the formal structure of the cost functions in conjunction with the difference in behaviour close to QS suggests that the different measures of QS will behave quite differently as optimisation cost functions. \par
From a formal perspective, we would expect to find $f_B$ and $f_C$ to perform similarly, while the non-linear $f_T$ to differ significantly. We now present some examples to illustrate the differences in the optimisation. We consider an example optimisation problem in which QS is targeted throughout the plasma volume. The optimisation parameters correspond to the Fourier modes describing the outermost surface, 
\begin{equation}
\begin{aligned}
    R &= \sum_{nm} R_{nm} \cos(m \vartheta - n N_p \phi_c) \\
    Z &= \sum_{nm} Z_{nm} \sin(m \vartheta -  n N_p \phi_c),
\end{aligned}
\label{eq:surface_harmonics}
\end{equation}
where $N_p$ is the number of field periods, $\phi_c$ is the cylindrical angle, and $\vartheta$ is a general poloidal angle.  The configuration optimised is based on a quasi-helically symmetric configuration\footnote{The basis of the optimisation is the \texttt{34DOF\_varyBoundary\_targetQuasisymmetry} scenario in https://github.com/landreman/stellopt\_scenarios.} with $N_p=4$ field periods. We shall investigate two aspects of the problem. First, we consider slices of parameter space with the purpose of observing the differences in QS measures in practice beyond the region close to the minima. This should give a partial idea of how the problem changes as a result of changing the form of the optimisation cost function. Then, we perform various optimisations using STELLOPT (\cite{Lazerson2020}) with implementations of the three different forms of the cost functions. \par
\begin{figure*}
    \centering
    \includegraphics[width=0.9\textwidth]{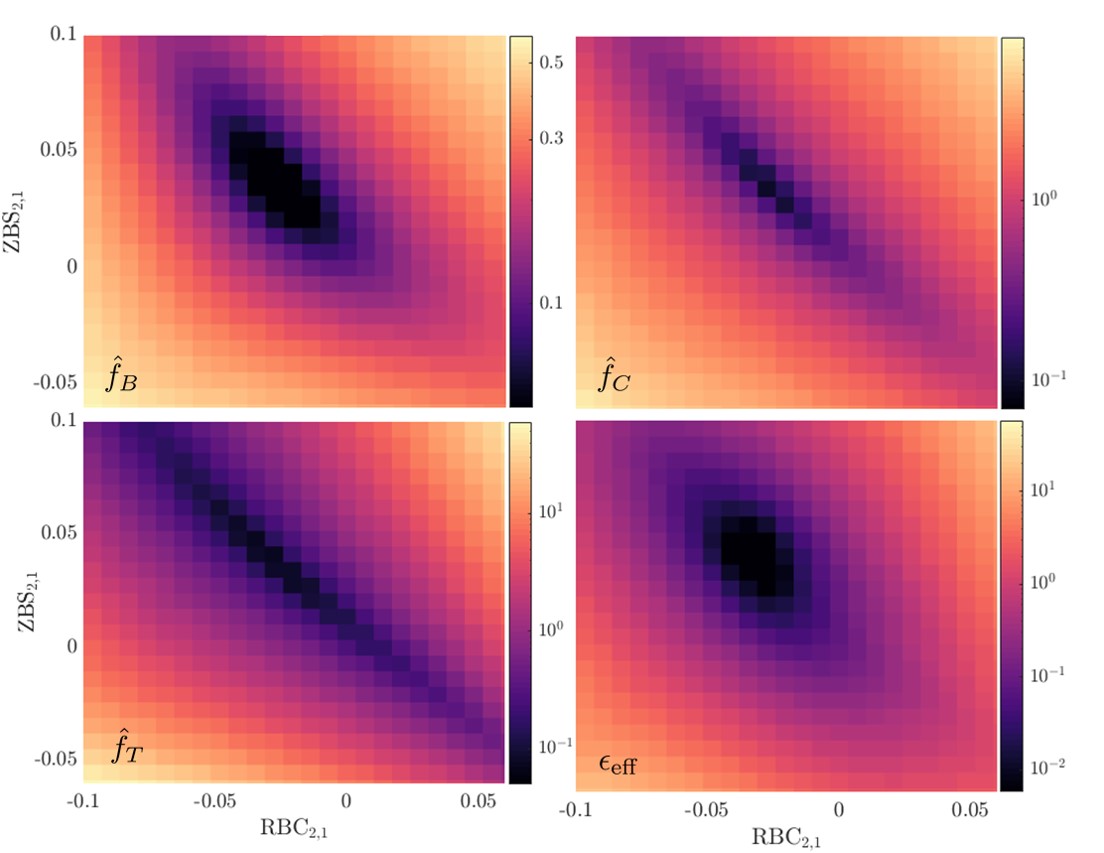}
    \caption{\textbf{Parameter space spanned by $n=2,m=1$ surface mode for $\hat{f}_B$, $\hat{f}_C$, $\hat{f}_T$ and $\epsilon_\mathrm{eff}$.} The plots show, clockwise, the cost functions $\hat{f}_B$, $\hat{f}_C$, $\epsilon_\mathrm{eff}$ and $\hat{f}_T$ as a function of the $(2,1)$ surface modes. The parameter scan is performed using the map option of STELLOPT around the stellarator design in the footnote of Page 15. The minima for each of these is found (clockwise) at: $(-0.031,0.031)$, $(-0.031,0.031)$, $(-0.037,0.045)$ and $(-0.037,0.037)$.}
    \label{fig:paramSpace}
\end{figure*}
Let us start with the exploration of parameter space. Figure \ref{fig:paramSpace} presents the evaluation of the QS metrics and $\epsilon_\mathrm{eff}$ as a function of two parameters characterising the surface (the $n=2,~m=1$ modes of the boundary \eqref{eq:surface_harmonics}). 
The metrics are evaluated at six equally-spaced magnetic flux surfaces, and their contributions are added together.\par
The objective landscape varies depending on the metrics used. Qualitatively, though, all the measures show a similar behaviour. The metrics share a region of reduced cost, which seem to roughly coincide. In a global sense, the metrics seem not to disagree with each other. This might appear as a surprise given the observations of the QS configurations and some of the formal discussion above. We should, however, remember that there exist certain changes that affect all metrics favourably. Away from a QS configuration it thus seems that there are directions in which one may reduce many of these modes such that all metrics decrease. This is supported by the parameter space representations in Fig.~\ref{fig:paramSpace}. However, note that the similarity in global trends does not preclude differences in local trends. \par
First, the relative variation of the cost functions in space is significantly different. Gradients seem to be relatively strong for the highly non-linear measures $\hat{f}_T$ and $\epsilon_\mathrm{eff}$, while the other two metrics behave more smoothly. This distinction in the gradients also seems to leave a region close to the minima much shallower for $\hat{f}_T$ and $\epsilon_\mathrm{eff}$, comparatively, which may lead to difficulties in accessing the exact minima. Secondly, beyond these differences in magnitude, there are also differences in the shape and location of the minima. The region near a minimum can be more or less elongated depending on the measure, and importantly, the minimiser is located at different values. In this particular case, minimisers vary by up to $\sim 20-50$\% in the boundary coefficients between different measures (although $\hat{f}_B$ and $\hat{f}_C$ have, in this case, roughly the same minimiser). In brief, although far from a minimum the various metrics may exhibit a qualitatively similar behaviour, close to the minimum there are significant differences. These latter differences are consistent with the discussion on the lack of universality in Sec. \ref{sec:universality}. \par
Of course, this illustration of the optimisation space is only one slice of a multidimensional manifold. The large number of degrees of freedom means that the parameter space slices will change as other parameters do. 
In general, this makes optimisation a highly complex process in which the interaction between different dimensions may amplify the differences between metrics. In addition, the algorithm employed to move around in optimisation space will also affect the result of a given optimisation. A discussion on how the optimisation is likely to perform is therefore extremely difficult, and we instead present an illustrative example. With that being said, from our discussion we expect the optimisation of $\hat{f}_B$ and $\hat{f}_C$ to be similar, while the non-linearity of $\hat{f}_T$ to change the optimisation significantly. We hypothesise here that there is a rough relation between changes in boundary shape and the Fourier content of $B$.  \par
\begin{figure}
    \centering
    \includegraphics[width=0.7\textwidth]{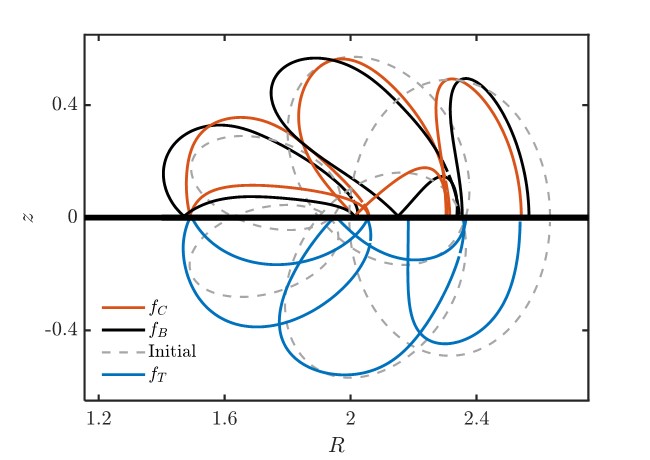}
    \caption{\textbf{Comparison of cross sections of optimisation results for different cost functions.} Plot showing the cross section of the stellarators as obtained from the optimisation using different forms of the cost function. The optimisation stops when the optimiser is unable to make further improvement (the characteristic change in the parameters during an iteration is $10^{-4}-10^{-3}$). 
    The broken line represents the starting point, while the black, blue and red lines represent the $f_B$, $f_T$ and $f_C$ optimisation results respectively. The cross sections are only presented in one half of the plane but the missing plane can be reconstructed symmetrically. }
    \label{fig:optCross}
\end{figure}

We present the resulting cross sections and values associated with the various metrics from optimisations in which QS is targeted in a volume using the different forms of QS metrics as cost functions in Fig.~\ref{fig:optCross} and Tab.~\ref{tab:optProps}. We start from a configuration that is deliberately chosen to be some distance away in parameter space from the QS configuration found in the footnote of Page 15. The optimisation uses a BFGS gradient-based method (as part of the suite MANGO\footnote{\label{foot:git_mango} https://github.com/hiddenSymmetries/mango.}) in which there is a single scalarised cost function with 32 degrees of freedom in the form of Fourier components (from 0 through 3, keeping the major radius $R_{00}$ fixed) defining the surface as in \eqref{eq:surface_harmonics}. The process does not penalise any additional property such as the commonly-targeted rotational transform or aspect ratio. By studying a simplified problem, we hope to gain more insight on the differences between the various QS metrics. The optimisation is performed for several values of the finite-difference step size to ensure convergence. \par

Comparing cross-sections in Fig.~\ref{fig:optCross} and the values in Tab.~\ref{tab:optProps} we observe that the resulting optima are quite different from one another. This emphasises the points discussed in this paper: that the precise form of the cost function has important qualitative implications. The optimisation using the standard form of $\hat{f}_B$ seems, in this case, to perform best in allowing the configuration to reach a state of minimal transport and QS measures. The optimisation by $\hat{f}_C$ is not far away, and a comparison of the cross sections shows similar shaping features. Interestingly, optimising for $\hat{f}_C$ does not seem to yield the smallest value for $\hat{f}_C$ (which $\hat{f}_B$ optimisation does). This suggests that the optimisation for $\hat{f}_C$ gets stuck in some effective local minimum. \par
In light of the comparison made in Sec. \ref{sec:QScomparison}, and with the hypothesis that the surface variations are roughly related to variations in $B$, we can think of optimisation as a process that rearranges energy in the space of Fourier modes of $|\mathbf{B}|$. Thinking in these terms may shed some light on understanding the differences in optimisation. In this rearrangement of energy in Fourier space, $\hat{f}_B$ contains a large space of degenerate interchanges. Energy can be freely exchanged between asymmetric modes of $B$, regardless of their mode numbers. In that sense, reducing $\hat{f}_B$ means, necessarily, reducing some asymmetric mode. On the other hand, the cost function $\hat{f}_C$ distinguishes between modes. Different weights at different mode numbers lifts part of the degeneracy in the exchange of mode-energy that occurs for $\hat{f}_B$. This distinction effectively excludes certain regions of the mode space from being used for the restacking of the mode-energy. In addition to decreasing the magnitude of an asymmetric mode, moving energy to modes closer to the symmetry direction will decrease $\hat{f}_C$. This could be a mixed blessing. Having this additional mechanism could be a way to avoid certain local minima by adding a new direction of descent. However, it could also present additional local minima if the optimisation prematurely reduces modes very close to the symmetry direction.
\par 
Relating changes in the mode content to changes in the boundary, we expect these two metrics to facilitate different ways of modifying the boundary. It is not clear which metric will lead to improved optimisation. We have here presented a single numerical exercise in which $\hat{f}_B$ appeared to perform better, but the performance of these metrics will likely depend on the particulars of the optimisation problem as well as other aspects such as the optimisation algorithm used.
\par
The optimisation of $\hat{f}_T$ in the presented example exhibits the worst performance of all. The cross sections show large differences with respect to the other configurations, suggesting that optimisation has stopped at a distinct local minimum. 
That the measure $\hat{f}_T$ is lower in the $\hat{f}_B$-optimised case serves as evidence that the optimisation of $\hat{f}_T$ teriminated at a different local minimum. The presence of nonlinearities in $f_T$ are perhaps responsible for this additional minimum. In the light of the Fourier space energy picture, $\hat{f}_T$ not only depends on the magnitude and the location of the mode energy, but also on the relative location of modes. This nonlinear coupling complicates the problem, potentially leading to additional local minima. Furthermore, the nonlinearity of $\hat{f}_T$ results in an objective landscape which appears flatter near the minimum (see Fig.~\ref{fig:paramSpace}) and may difficult attaining an accurate optimum. Although this qualitative picture fits with observations made for this optimisation example, we remark that the behaviour is likely to vary between problems. 

\par
In the example presented, the aspect ratio of the configuration does not seem to change significantly between and along optimisations (values lie within $\sim15\%$ of the initial point). This seems counterintuitive, as from a near-axis perspective one would expect a larger aspect ratio to allow for a more quasisymmetric solution. This follows from the possibility of satisfying QS conditions sufficiently close to the magnetic axis, as the relevant expansion parameter for near-axis scales as $\sqrt{\psi/B}/R$ (\cite{landreman2019}). Since we are fixing the total flux, decreasing the minor radius of the toroidal configuration can increase the differential flux over which the QS error is small. In practice, however, the optimiser has difficulty in finding these large-aspect-ratio minima. The difficulty to change the aspect ratio without sacrificing QS along the way and the choice of parameterisation (\cite{Henneberg2021}) \eqref{eq:surface_harmonics} might provide an explanation for this behaviour. \par
\par
\begin{table}
    \centering
    \begin{tabular}{c|c|c|c|c|c|}
        Opt. & $\hat{f}_B$ & $\hat{f}_C$ & $\hat{f}_T$ & $\epsilon_\mathrm{eff}$ & Asp.  \\\hline\hline
        Start & 0.54 & 5.3 & 24 & 66 & 6.3 \\
        $f_B$ & 3.3e-3 & 2.4e-2 & 3.0e-2 & 5.1e-5 & 7.1 \\
        $f_C$ & 2.3e-2 & 7.6e-2 & 0.50 & 3.9e-3 & 6.8 \\
        $f_T$ & 6.2e-2 & 1.9e-1 & 4.8e-2 & 3.6e-2 & 6.0
    \end{tabular}
    \caption{\textbf{Averaged metric values for optimised stellarators.} The table summarises the average value of the metric over the volume of the optimised stellarators with respect to the different cost functions.}
    \label{tab:optProps}
\end{table}
What can we thus draw from this illustrative example? First, the form of the cost function must be carefully considered. Here we only examined QS and observed that the optimisation result varies significantly depending on the form of the cost function. Similar behaviour will likely affect the optimisation of other physical measures. Second, we see in this example that the `simplest' forms (in this case, $f_B$ and $f_C$) seemed to provide the better global performance. (By simpler we mean a lower number of field derivatives and a simpler dependence on the Fourier modes.) Increased complexity seems to make optimisation harder. However, it is difficult to make general statements based only on this example, and there might be exceptions not treated in this paper. \par
As discussed in the previous section, close to local QS minima, improvements in QS metrics do not always lead to improvements in physical properties. Thus insisting on making marginal improvements on $\hat{f}_B$ (and for that matter, any of the other QS measures) does not seem to be a useful exercise. Close to a QS minima it does seem appropriate to prioritise optimisation with respect to more physical measures. Perhaps insisting on QS at all times may stand its ground if an absolute QS zero could be reached (a point with clear physical meaning), such as in the optimisation for QS on a single magnetic surface. But even then, stellarator design usually requires additional physical considerations.\par We seem then to need to alternate between an optimisation dominated by the QS metrics and an optimisation guided by other physics metrics. A way forward is to impose QS as an inequality constraint, so that physical metrics can dominate optimisation when QS is `small enough.' Of course there is the unavoidable difficulty of choosing the value of a QS metric to indicate `sufficient QS.' Picking a fixed value will modify the shape of optimisation space in a non-trivial way. Instead of choosing a fixed value, a more dynamic alternative would be to turn QS optimisation on and off guided by the relative changes or gradient norms of the QS metrics. One might choose to not optimise for QS at all, instead initialising the search from an already fairly quasisymmetric configuration. This would place the optimisation in the right basin, but lead to refinements in it based on physically-relevant features. Possible initial seeds could be constructed from near-axis solutions. \par
As explored, the non-universality of the QS metrics close to minima seems to require amendment by introducing additional physics metrics. At the same time, detailed physical calculations are likely to lead to complicated parameter spaces including multiple local minima compared to those of the simpler QS metrics. Using QS metrics in conjunction with a more complex physical metric may provide the best of both worlds. This could shed some light on recent results in (\cite{bader2019}). In this work, optimisation of QS through $f_B$ together with a more complex physical proxy for $\alpha$ particle losses seems to perform better than each of the pieces separately. We hypothesise that $f_B$ allows for a smoother global navigation (avoiding some potential local minima), while the physics metric helps in finding a more physically meaningful final minimum. Using both metrics in an appropriately balanced way seems to lead to a convenient compromise in which benefits of both aspects are brought together. \par
Stellarator design often requires multi-objective optimisation. Given the nonlinearity of the objectives of interest, adding additional terms to the cost function can change the problem dramatically. Thus, the differences observed above for the QS metrics will only be more complex in the presence of other physics objectives. Those differences may be amplified or reduced depending on their relation to the behaviour of the other metrics. Such an analysis will be necessary to reason how to best compose QS with these other metrics. In these complex scenarios, the idea of initially seeding optimisation with an approximately QS configuration gains strength, as it would make the problem simplest. Further study is required to verify these proposed optimisation strategies.

\section{\label{sec:conclusion} Conclusion}
In this paper we have explored three main measures of quasisymmetry: the standard Fourier form, the two-term formulation and the triple product form. We performed an analytical comparison of their origin and mathematical structure. Particularly important are their differences in weighting the asymmetric modes of $B$ in Boozer coordinates. While the standard Fourier form treats all modes equally, the two-term form weights them differently, and the triple vector formulation involves a mode convolution.  \par
The differences in these valid QS metrics shows that there is no universal measure of QS. While the measures are all rigorously equivalent when QS is satisfied exactly (assuming $\textbf{j} \cdot \nabla \psi = 0$), there is no unique measure that quantifies deviations from QS. This lack of universality makes the comparison between configurations particularly difficult near approximately QS solutions, and some physically relevant property (such as neoclassical transport through $\epsilon_\mathrm{eff}$ or $\alpha$ particle confinement) would be needed for a more meaningful comparison. Near those local minima that are not exactly QS, optimising QS cost functions for marginal improvements generally lacks physical meaning. Away from the minima, the global structures of the QS metrics are qualitatively similar, providing more physical meaning to the metrics. \par
We also show that the mathematical form of the measure affects the optimisation significantly via an example. In the particular example presented, the complexity of a cost function seems to lead to ending in alternative local minima.\par
In light of the above, we suggest either enforcing approximate QS through an inequality constraint (so that optimisation close to minima is dominated by some direct physics measure) or through an appropriate initial configuration. Additional attention should also be paid to the form of physics or engineering cost functions.

\section*{Acknowledgements}
The authors would like to acknowledge Daniel Dudt for fruitful discussion. This research is primarily supported by a grant from the Simons Foundation/SFARI (560651, AB) and DoE Contract No DE-AC02-09CH11466.

\section*{Data availability}
The data that support the findings of this study are available from the corresponding author upon reasonable request.

\appendix

\section{Magnetic differential relation between $f_C$ and $f_T$}

Although the triple vector product form of QS seems distinct from the other forms given its non-linear character, we can establish a simple relation between $f_T$ and $f_C$.
To obtain this relation it is most convenient to use the same tools we used in the derivations of the different forms of QS in Sec.~2. Away from QS, one may still construct the symmetry field as $\tilde{\mathbf{u}}=\nabla\Phi\times\nabla B/(\mathbf{B}\cdot\nabla B)$, where we assume $\mathbf{B}\cdot\nabla B\neq0$. At extremal points of the magnetic field magnitude along $\mathbf{B}$, this construction fails. 
We have seen the role of the extrema, for example, in the discussion of the single-particle dynamics (Sec. \ref{sec:QScomparison}). Requiring pseudo-symmetry takes care of these, as it guarantees $\mathbf{B}\cdot\tilde{\mathbf{u}}$ to be well-behaved. Constructing this vector field $\tilde{\mathbf{u}}$ in this way is equivalent to enforcing (\ref{eqn:C1}) and (\ref{eqn:C2}). With these as well as $\nabla\cdot\mathbf{B}=0$, 
\begin{equation*}
    \mathbf{B}\cdot\nabla(\mathbf{B}\cdot\tilde{\mathbf{u}})=B^2\nabla\cdot\tilde{\mathbf{u}},
\end{equation*}
using $\mathbf{j}\cdot\nabla\psi=0$ and appropriate vector identities to rewrite the LHS. Now, from $\mathbf{B}\cdot\tilde{\mathbf{u}}/\Phi'=C+f_C/(\mathbf{B}\cdot\nabla B)$, it then follows that $f_C$ and $f_T$ are related to each other through a magnetic differential equation (MDE),
\begin{equation}
    \mathbf{B}\cdot\nabla \left(\frac{f_C}{\mathbf{B}\cdot\nabla B}\right)=-\frac{f_T}{(\mathbf{B}\cdot\nabla B)^2}. \label{eqn:fT2fP}
\end{equation}
This MDE shows that $f_T$ does, in fact, contain a higher order derivative in comparison to $f_C$.

\bibliographystyle{jpp}
\bibliography{costFunctionQS}

\end{document}